\documentclass[%
twocolumn,
aps,authoryear
]{aastex63}
\usepackage{graphicx}
\usepackage{epstopdf}
\usepackage{ulem}
\usepackage{bm}
\usepackage{color}
\usepackage{textcomp}
\usepackage{gensymb}
\usepackage{hyperref}
\usepackage{amssymb,amsmath}

\newcommand{\isois}{IS\(\odot\)IS}

\newcommand{\R}{\hbox{$\cal R$}}

\renewcommand{\S}{\hbox{$\cal S$}}

\newcommand{\M}{\hbox{$\cal M$}}
\newcommand{\E}{\hbox{$\mathcal{E}$}}

\begin{document}
\title{Clustering of Intermittent Magnetic and Flow Structures near Parker Solar Probe's First Perihelion -- A Partial-Variance-of-Increments Analysis}
\author[0000-0002-7174-6948]{Rohit Chhiber}
\correspondingauthor{Rohit Chhiber}
\email{rohitc@udel.edu}
\affiliation{Department of Physics and Astronomy, University of Delaware, Newark, DE, USA}
\affiliation{Code 671, NASA Goddard Space Flight Center, Greenbelt, MD, USA}
\author[0000-0002-5317-988X]{M L.~Goldstein}
\affiliation{Code 672, NASA Goddard Space Flight Center, Greenbelt, MD, USA}
\affiliation{University of Maryland Baltimore County, Baltimore, MD, USA}
\author[0000-0002-2229-5618]{B. A.~Maruca}
\affiliation{Department of Physics and Astronomy, University of Delaware, Newark, DE, USA}
\affiliation{Bartol Research Institute, University of Delaware, Newark, DE 19716, USA}
\author[0000-0001-8478-5797]{A. Chasapis}
\affiliation{Department of Physics and Astronomy, University of Delaware, Newark, DE, USA}
\affiliation{Laboratory for Atmospheric and Space Physics, University of Colorado, Boulder, CO, USA}
\author[0000-0001-7224-6024]{W. H. Matthaeus}
\affiliation{Department of Physics and Astronomy, University of Delaware, Newark, DE, USA}
\affiliation{Bartol Research Institute, University of Delaware, Newark, DE 19716, USA}
\author[0000-0003-3414-9666]{D. Ruffolo}
\affiliation{Department of Physics, Faculty of Science, Mahidol University, Bangkok, Thailand}
\author[0000-0002-6962-0959]{R. Bandyopadhyay}
\affiliation{Department of Physics and Astronomy, University of Delaware, Newark, DE, USA}
\author[0000-0003-0602-8381]{T. N. Parashar}
\affiliation{Department of Physics and Astronomy, University of Delaware, Newark, DE, USA}
\author[0000-0001-8358-0482]{R. Qudsi}
\affiliation{Department of Physics and Astronomy, University of Delaware, Newark, DE, USA}

\author[0000-0002-4401-0943]{T. Dudok de Wit}
\affiliation{LPC2E, CNRS and University of Orl\'eans, Orl\'eans, France}

\author[0000-0002-1989-3596]{S. D. Bale}
\affiliation{Space Sciences Laboratory, University of California, Berkeley, CA, USA}
\affiliation{Physics Department, University of California, Berkeley, CA, USA}
\affiliation{The Blackett Laboratory, Imperial College London, London, UK}

\author[0000-0002-0675-7907]{J.~W. Bonnell}
\affiliation{Space Sciences Laboratory, University of California, Berkeley, CA, USA}

\author[0000-0003-0420-3633]{K. Goetz}
\affiliation{School of Physics and Astronomy, University of Minnesota, Minneapolis, MN, USA}

\author[0000-0002-6938-0166]{P.~R. Harvey}
\affiliation{Space Sciences Laboratory, University of California, Berkeley, CA, USA}

\author[0000-0003-3112-4201]{R.~J. MacDowall}
\affiliation{Code 695, NASA Goddard Space Flight Center, Greenbelt, MD, USA}

\author[0000-0003-1191-1558]{D. Malaspina}
\affiliation{Laboratory for Atmospheric and Space Physics, University of Colorado, Boulder, CO, USA}

\author[0000-0002-1573-7457]{M. Pulupa}
\affiliation{Space Sciences Laboratory, University of California, Berkeley, CA, USA}

\author[0000-0002-7077-930X]{J.~C. Kasper}
\affiliation{Climate and Space Sciences and Engineering, University of Michigan, Ann Arbor, MI, USA}
\affiliation{Smithsonian Astrophysical Observatory, Cambridge, MA, USA}

\author[0000-0001-6095-2490]{K.~E. Korreck}
\affiliation{Smithsonian Astrophysical Observatory, Cambridge, MA, USA}

\author[0000-0002-3520-4041]{A.~W. Case}
\affiliation{Smithsonian Astrophysical Observatory, Cambridge, MA, USA}

\author[0000-0002-7728-0085]{M. Stevens}
\affiliation{Smithsonian Astrophysical Observatory, Cambridge, MA, USA}

\author[0000-0002-7287-5098]{P. Whittlesey}
\affiliation{Space Sciences Laboratory, University of California, Berkeley, CA, USA}

\author{D. Larson}
\affiliation{Space Sciences Laboratory, University of California, Berkeley, CA, USA}

\author[0000-0002-0396-0547]{R. Livi}
\affiliation{Space Sciences Laboratory, University of California, Berkeley, CA, USA}

\author{M. Velli}
\affiliation{Department of Earth, Planetary, and Space Sciences, University of California, Los Angeles, CA, USA}

\author[0000-0003-2409-3742]{N. Raouafi}
\affiliation{Johns Hopkins University Applied Physics Laboratory, Laurel, MD, USA}

\begin{abstract}
During the Parker Solar Probe's (PSP) first perihelion pass, the spacecraft reached within a heliocentric distance of \(\sim 37~R_\odot\) and observed numerous magnetic and flow structures characterized by sharp gradients. To better understand these intermittent structures in the young solar wind, an important property to examine is their degree of correlation in time and space. To this end, we use the well-tested Partial Variance of Increments (PVI) technique to identify intermittent events in FIELDS and SWEAP observations of magnetic and proton-velocity fields (respectively) during PSP's first solar encounter, when the spacecraft was within 0.25 au from the Sun. We then examine distributions of waiting times between events with varying separation and PVI thresholds. We find power-law distributions for waiting times shorter than a characteristic scale comparable to the correlation time, suggesting a high degree of correlation that may originate in a clustering process. Waiting times longer than this characteristic time are better described by an exponential, suggesting a random memory-less Poisson process at play. These findings are consistent with near-Earth observations of solar wind turbulence. The present study complements the one by Dudok de Wit et al. (2019, present volume), which focuses on waiting times between observed ``switchbacks'' in the radial magnetic field.
\end{abstract}
%
%
%
\section{Introduction}\label{sec:intro}
Turbulence has diverse 
effects in fluids, magnetofluids, and plasmas such as the interplanetary medium and the solar wind \citep{pope2000book,
biskamp2003magnetohydrodynamic,matthaeus2011SSR}. 
While far less is understood about the latter case compared to the two fluid cases, plasma turbulence 
apparently shares with classical turbulence 
its capacity to greatly enhance transport. 
This includes the transport of energy across scales, 
suggested by the presence of characteristic 
second-order statistics such as wavenumber 
spectra \citep{bruno2013LRSP}, as well as third-order statistics \citep{politano1998PRE} which
quantitatively \citep[and in some cases, approximately; see][]{hellinger2018ApJ} characterize the rate of cascade across scales.
Turbulence also produces intermittency 
\citep{sreenivasan1997AnRFM,matthaeus2015ptrs},
and the associated coherent structures, including 
current sheets and vortices \citep{zhdankin2012PRL,parashar2016ApJ}, 
are responsible for 
spatial concentration
of physical processes observed in heliospheric plasmas, 
such as heating, heat conduction, temperature anisotropies, 
and local particle energization  \citep{osman2011ApJ,greco2012PRE,karimabadi2013coherent,tessein2013ApJ}.
Coherent current and field structures can also 
significantly influence the transport of field lines and 
charged particles, affecting 
distributed transport and acceleration \citep{ruffolo2003ApJ,zank2014ApJ,tooprakai2016ApJ}.
There are, therefore, numerous 
applications that provide motivation for 
better understanding of the occurrence rate and distribution 
of intensities of 
coherent structures such as current sheets.

Traditionally, sharp changes 
in the magnetic field have been classified as various type of ``discontinuities'', 
which are convected or propagated 
as approximate solutions of linear ideal magnetohydrodynamics (MHD)
\citep{burlaga1969SoPh,tsurutani1979JGR,neugebauer1984JGR,
neugebauer2010SWconf}. The ongoing recognition that these 
structures are generated rapidly and generically 
by turbulence changes the nature of their study at a level of basic physics. 
As a consequence of turbulence,
coherent structures 
are a manifestation of nonlinear dynamics and intermittency, 
a direct reflection of the higher-order 
correlations that are implied by 
the cascade process itself
\citep{oboukhov1962JFM,frisch1995book}. Such higher-order statistical correlations are routinely  
observed in space observations in 
plasmas such as the solar wind 
\citep{horbury1997NPGeo,sorriso-valvo1999GRL,chhiber2018MMS}.

For these reasons, as Parker Solar Probe \citep[PSP;][]{fox2016SSR}
explores regions of the heliosphere previously inaccessible 
to in-situ
observation, several baseline questions arise concerning the 
observed coherent structures, 
whether one chooses to call them 
discontinuities, 
structures, or current sheets and vortex sheets. Since these 
coherent structures 
are routinely observed at 1 au and elsewhere 
and are often found to be related to flux-tube structures, \citep{borovsky2008JGR,neugebauer2015JGRA,zheng2018ApJ,pecora2019ApJ}, a description of their 
statistical distribution 
along PSP's orbits becomes a matter 
of theoretical interest as well as 
considerable import in 
the various 
heliospheric plasma physics applications 
alluded to above. 
Here we make use of a 
simple and well-studied method for characterizing 
statistics of 
coherent structures, 
namely the Partial Variance of Increments 
(PVI) method \citep{greco2009ApJ,greco2018ssr}, and apply it to characterize coherent magnetic field 
and velocity field structures during the first PSP solar encounter. 

The paper is organized as follows -- Section \ref{sec:backg} defines the PVI measure and provides some background; in Section \ref{sec:data} we describe the data used and its processing; Sections \ref{sec:mag} and \ref{sec:vel} describe the results of the analyses of the magnetic and velocity PVI, respectively; we conclude with a discussion in Section \ref{sec:conclude}; Appendix \ref{sec:app} discusses the association between power-law waiting times and processes that can be described as a Cantor set.

\section{Background}\label{sec:backg}
The Partial Variance of Increments (PVI) is a well-tested measure of the occurrence of sharp gradients in the magnetic field -- discontinuities, 
current sheets, sites of magnetic reconnection, etc. See \cite{greco2018ssr} for a review of applications. Such discontinuities are believed to play a key role in enhanced dissipation, and particle heating \citep[e.g.,][]{chasapis2015ApJ} and energization \citep[e.g.,][]{tessein2013ApJ} in space plasmas.
If we consider a given time lag between measurements, for lags much larger than the correlation time, measurements of turbulent velocity and magnetic fields are typically uncorrelated and distributions of increments are Gaussian. However, for time lags corresponding to distances within the inertial range, probability density functions (PDFs) of increments have ``fatter'' non-Gaussian tails and are fit better by stretched exponential, lognormal, truncated L\'evy-flight, and kappa distributions \citep{kailasnath1992PRL,sorriso-valvo1999GRL,bruno2004EPL,pollock2018JASTP}. Distributions of waiting times between high PVI ``events'' typically exhibit power laws at inertial range lags and exponential behavior at longer, uncorrelated lags \cite{greco2009PRE,greco2018ssr}. Power-law behavior indicates a correlated ``clustering'' process which is statistically self-similar in time and possesses ``memory'', as opposed to a random Poisson process which results in exponential waiting-time distributions. Waiting-time analyses have been employed to make this distinction in diverse fields of study: space physics \citep{boffetta1999PRL,lepreti2001ApJ,DAmicis2006AnGeo,
greco2009PRE}, geophysics \citep{carbone2006prl}, laboratory materials \citep{ferjani200PRE}, and seismology \citep{mega2003prl}, to name a few. 

Other studies analyzing PSP data (this volume) have revealed
many sharp jumps 
in magnetic field measurements by FIELDS and proton velocity measurements by SWEAP during PSP's first solar encounter \citep{bale2019nature,Horbury2019psp}. 
Dudok de Wit et al. (2019)  
have examined statistical 
distributions of events 
identified by inspection
as ``switchbacks'' or ``jets''. Examination of the same data using the PVI method represents a complementary approach, since the PVI is not tailored to a specific type of discontinuity but is instead a general tool for identifying a broad class of intermittent structures \citep{greco2018ssr}. Another motivation for this study lies in the fact that the PVI technique is used in two other concurrent studies submitted to the PSP Special Issue ApJ -- (1) \cite{Bandyopadhyay2019psp_pvi} examine the association of energetic-particle fluxes from  \isois~with intermittent magnetic structures, as identified by the PVI technique; (2) \cite{Qudsi2019psp} study the association of high proton-temperatures measured by SWEAP with high magnetic-PVI values. 

The PVI is essentially the magnitude of the (vector) increment in a field at a given lag, normalized by the variance of the field. Note that the increment of a turbulent field has long occupied a central role in turbulence research, with particular importance having been given to moments of the increment, the so-called structure functions \citep[e.g.,][]{Monin1971book,tu1995SSRv}. The PVI is related to the first-order structure function, but is distinct in that it is a pointwise (not-averaged) measure. For the magnetic field \(\bm{B}\) the PVI at time \(s\) is defined, for lag \(\tau\) in time, as \citep{greco2018ssr}:
\begin{equation}
\text{PVI}_{s,\tau} = |\Delta \bm{B}(s,\tau)|/\sqrt{\langle |\Delta \bm{B}(s,\tau)|^2\rangle}, \label{eq:pvi}
\end{equation}
where the \(\langle .\rangle\) denotes averaging over a suitable interval \citep[see][]{isaacs2015JGR120,jagarlamudi2019ApJ}. The increment is defined as \(\Delta \bm{B}(s,\tau) = \bm{B}(s+\tau) - \bm{B}(s)\). The velocity PVI is defined similarly. To compute the variance we use a moving average over a window ten times larger than the correlation time of magnetic or velocity fluctuations, as appropriate (see Sections \ref{sec:mag} and \ref{sec:vel}, below). Note that PVI captures gradients in each vector component of B. In the following figures, we will denote the magnetic and velocity PVI as \(\mathrm{PVI}_{\bm{B}}\) and \(\mathrm{PVI}_{\bm{V}}\), respectively.

Values of PVI \(>3\) have been associated with non-Gaussian structures, PVI \(>6\) with current sheets, and PVI \(>8\) with reconnection sites \citep{greco2018ssr}. 
Events with PVI exceeding 3 
become progressively less likely 
to be a sample of a Gaussian
random process.
Therefore, even moderately large 
PVI selects samples
of intermittency, meaning, in this context, a sample taken 
from the ``fat tail'' portion
of a distribution
associated with a 
process that admits an 
elevated likelihood of extreme events \citep{sreenivasan1999RMP,matthaeus2015ptrs}.
\added{Note that the PVI method is one amongst several that have been developed for identifying discontinuities in turbulent flows, such as the wavelet-based Local Intermittency Measure \citep{veltri1999SolWind9,farge2001prl} and the Phase Coherence Index \citep{hada2003SSR}. See \cite{greco2018ssr} for a comparison of some of these methods with the PVI technique.}

\section{Data}\label{sec:data}
\begin{figure*}
\centering
\includegraphics[width=1\textwidth]{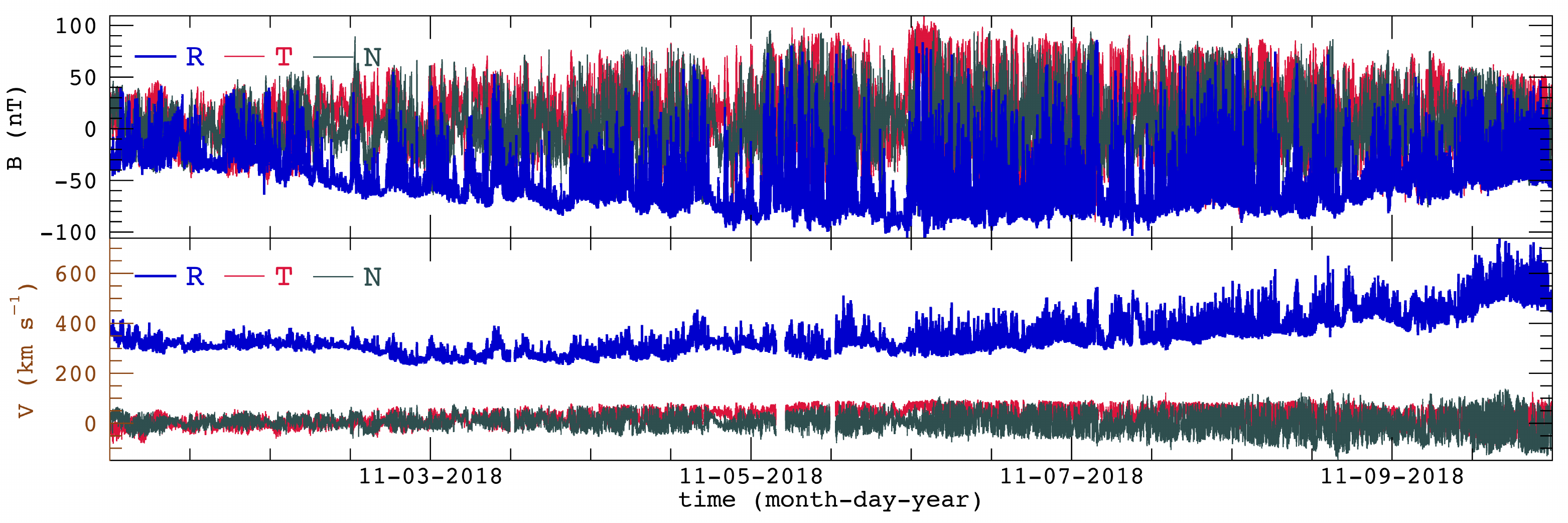}\caption{Time series of the heliocentric RTN components (blue, red, and green curves, respectively) of the magnetic and velocity fields during the period considered here are shown in the top and bottom panels, respectively. Note that the R-component is plotted using a thicker curve than the other two components.}
\label{fig:tseries_B_V}
\end{figure*}

We use magnetic-field data from the flux-gate magnetometer (MAG) aboard the FIELDS instrument suite \citep{bale2016SSR} and proton-velocity data from the Solar Probe Cup (SPC) on the SWEAP instrument suite \citep{kasper2016SSR,Case2019psp}, covering a period of about 10 days centered on the first perihelion. The magnetic field data used span the full range from UTC time 2018-11-01T00:00:00 to  2018-11-09T23:59:59, and have been resampled to 1-second cadence using linear interpolation. Note that data gaps are not an issue in MAG measurements during the period considered here. \added{The resampled magnetic data in heliocentric RTN coordinates \citep{franz2002pss} are shown in the top panel of Figure \ref{fig:tseries_B_V}. For a detailed description of these observations, including the large ``switchbacks'' in the radial magnetic field, see other papers in the present volume \citep[][Dudok de Wit et al. 2019]{bale2019nature,Horbury2019psp}.}

We use proton velocity measurements at 0.87-second resolution from the SPC, which are then processed to remove spurious or artificial spikes. \added{These data are shown in heliocentric RTN coordinates in the bottom panel of Figure \ref{fig:tseries_B_V}. Despite the numerous fluctuations, the bulk flow is fairly steady and well-described as slow wind (\(V_R < 500 ~\text{km}~\text{s}^{-1}\)) for most of the period considered here. During the last day PSP may have passed over a small coronal hole and sampled relatively fast wind above 600 \(\text{km}~\text{s}^{-1}\).} Data gaps are a more significant issue in SPC measurements during the first encounter (compared to MAG measurements), and we have used the following procedure to prepare the data for our analyses. We first split the time series of velocity measurements from 2018 November 1 to 2018 November 10 into 8-hour sub-intervals. We then discard sub-intervals that have data gaps larger than 10 seconds. The remaining sub-intervals have gaps with an average duration of about 1.5 s, and linear interpolation is used over these gaps. This procedure produces three periods -- (i) 2018-11-01T00:00:03 to 2018-11-03T08:00:02; (ii) 2018-11-05T16:00:03 to  2018-11-07T00:00:03; (iii) 2018-11-08T00:00:04 to 2018-11-10T08:00:03, within each of which we have continuous time series of velocity measurements at 0.87 s cadence. The PVI waiting-time analyses are performed separately within each of these three periods, and the results are then accumulated to obtain improved statistics (see details in Section \ref{sec:vel}, below). Note that reliable waiting-time estimation precludes the use of intervals with large data gaps. \added{Bulk plasma properties over the encounter are shown in Table \ref{tab:bulk}}.

\begin{table*}
\centering
  \begin{tabular}{| c | c | c | c | c | c | c | c | c |}
    \hline 
  Time & \(\langle V\rangle\) & \(\langle T_\text{i}\rangle\) & \(\langle n_\text{i}\rangle\) & \(d_\text{i}\) & \(\langle B\rangle\) & \(\langle \delta B\rangle\) & \(\langle V_\text{A}\rangle\) & \(\beta_\text{i}\)   \\ \hline    
      2018-11-01 to  2018-11-10 & \(350~ \text{km}~\text{s}^{-1}\) & \(0.6\times 10^6\) K & 215 \(\text{cm}^{-3}\) & 17 km & 70 nT & 50 nT & 100 \(\text{km}~\text{s}^{-1}\) &  1 \\ \hline
  \end{tabular}
\caption{Bulk plasma parameters for PSP's first solar encounter. Shown are average values of proton speed \(\langle V\rangle \equiv \langle \sqrt{V_R^2 + V_T^2 +V_N^2}\rangle\), proton temperature \(\langle T_\text{i}\rangle\), proton density \(\langle n_\text{i} \rangle\), proton inertial scale \(d_\text{i}\), magnetic field magnitude \(\langle B\rangle \equiv \langle \sqrt{B_R^2 + B_T^2 + B_N^2}\rangle\), the rms magnetic fluctuation \(\langle \delta B\rangle\equiv \sqrt{\langle |\bm{B} - \langle \bm{B}\rangle|^2 \rangle}\), Alfv\'en speed \(\langle V_\text{A}\rangle\), and proton beta \(\beta_\text{i}\). Averaging is performed over the entire duration of UTC time 2018-11-01T00:00:00 to 2018-11-10T23:59:59.}\label{tab:bulk}
\end{table*}

\section{PVI Analysis of the Magnetic Field}\label{sec:mag}

As mentioned in Section \ref{sec:backg}, to compute the PVI time-series we need estimates of the correlation time. We use the Blackman-Tukey technique  \citep[see][]{matthaeus1982JGR} with an averaging interval of 24 hours to compute the autocorrelation of the magnetic field. The correlation time is estimated as the time at which the autocorrelation falls to \(1/e\).  Note that the correlation time does not change significantly on using a 12-hour averaging interval instead of 24 hours. The paper by Parashar et al. in the present special issue shows correlation times computed in this way for the encounter. See also \cite{smith2001JGR}, \cite{isaacs2015JGR120}, \cite{jagarlamudi2019ApJ} and \cite{bandyopadhyay2019psp_cascade} (present volume) for discussions of subtleties and potential issues in accurate determination of correlation times. An alternative estimate of the correlation time may be obtained from the break frequency between the ``\(1/f\)'' and inertial ranges in the power spectrum of the magnetic fluctuations \citep{Chen2019psp}; we confirmed that this estimate is comparable to the correlation time we use here. Furthermore, the PVI does not appear to depend sensitively on the averaging interval used.

\begin{figure*}
\centering
\gridline{\fig{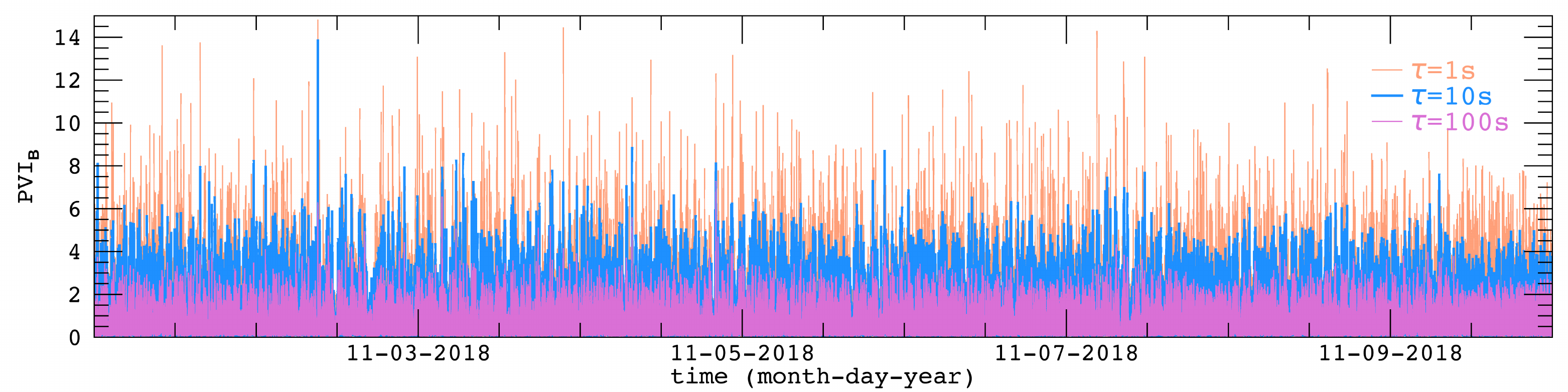}{1\textwidth}{(a)}}
\gridline{\fig{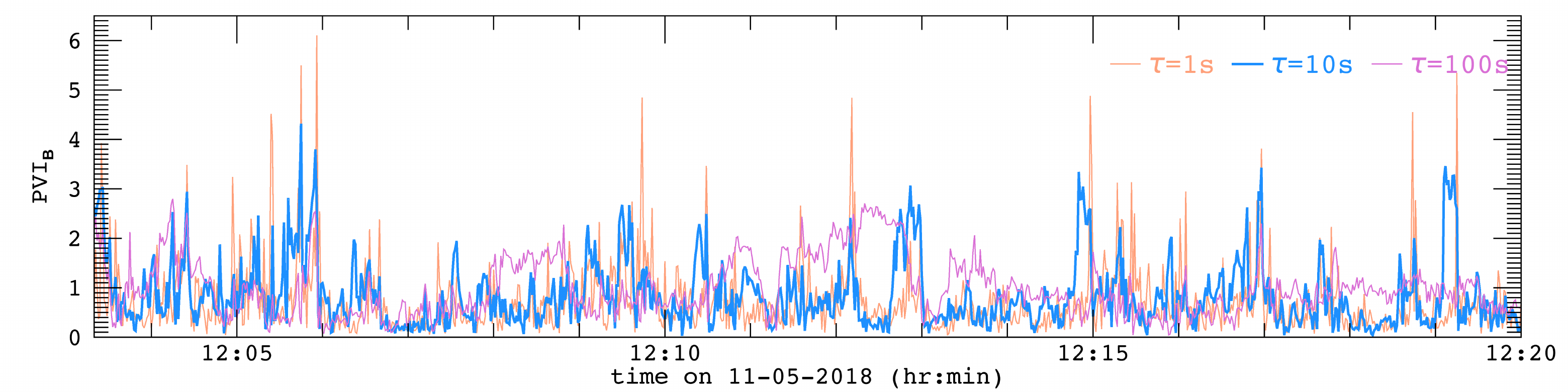}{1\textwidth}{(b)}}
\caption{(a) PVI (with lag \(\tau\) equal to 1, 10, and 100 s) time-series for magnetic field during the first encounter. (b) The same time series for about 15 minutes on 2018 November 5. In both panels the 10 s case is shown as a thicker line compared to the other two.}
\label{fig:pviB_time}
\end{figure*}

During the period analyzed here the magnetic correlation time varies from about 1000 s to 350 s. Accordingly, assuming an average correlation time of 600 s for the encounter, we use a rolling boxcar average over a window \(10\times 600\) seconds in duration to estimate the variance [the denominator in Equation \eqref{eq:pvi}] for the computation of the magnetic PVI. The resulting time series is shown in Figure \ref{fig:pviB_time} for three different lags: \(\tau = \) 1, 10, and 100 seconds. The 1 and 10 seconds lags lie well within the inertial range (the ion inertial scale corresponds to an approximate temporal lag of 0.05 s [Parashar et al. 2019, this issue]), while the 100 s lag is comparable to the correlation time. Note that as the lag \(\tau\) is increased we still sample over times with 1-second cadence. 

\begin{figure}
\centering
\includegraphics[width=.5\textwidth]{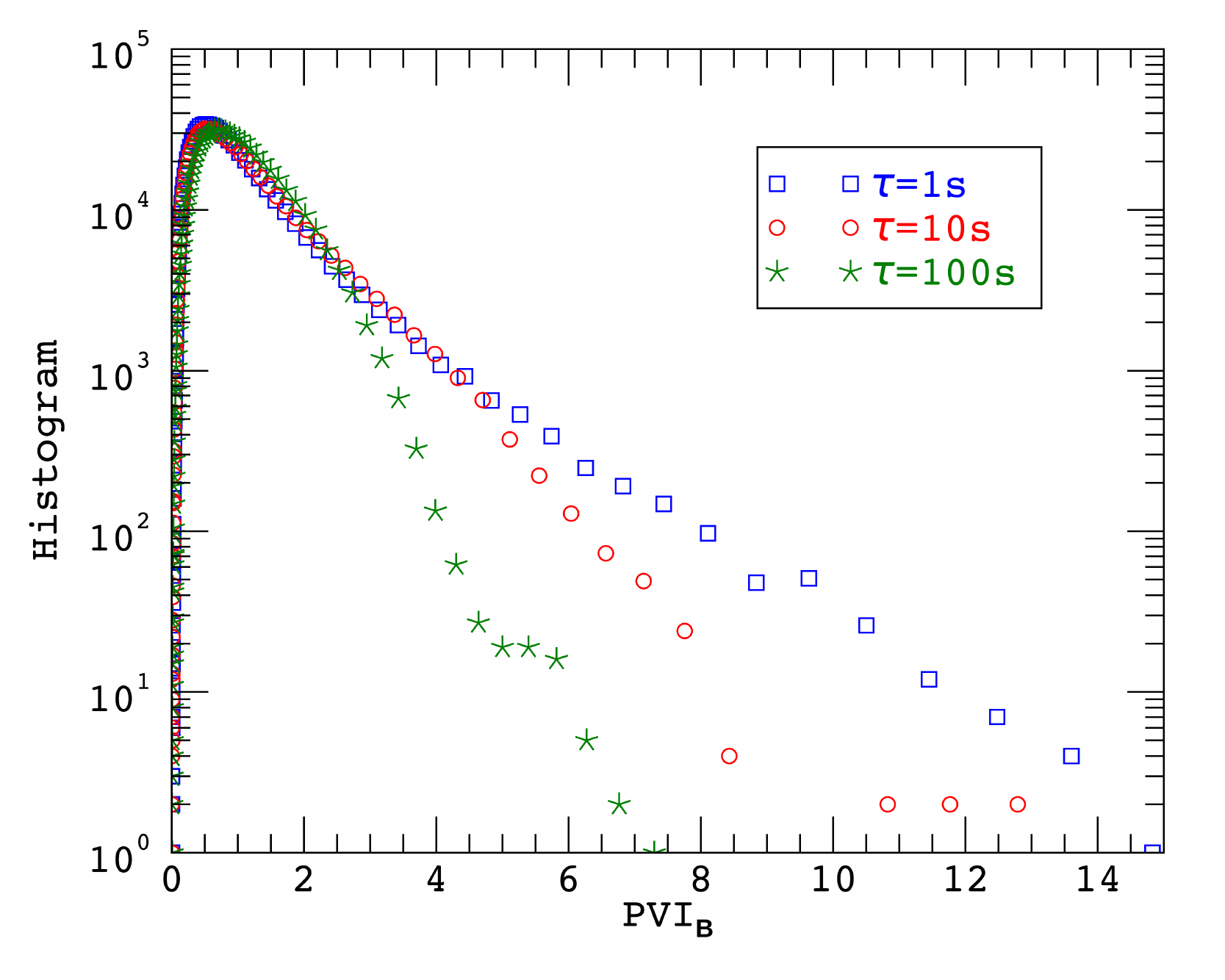}
\caption{Histograms (showing frequency of occurrence, or number of counts) of PVI values for different lags \(\tau\), for the magnetic field during the first encounter. Note the 
elevated likelihood of large PVI 
values at shorter lags, indicative 
of enhanced small-scale intermittency, typical of non-Gaussian processes and turbulence.
}
\label{fig:pvi_pdf}
\end{figure}

It is  clear from Figure \ref{fig:pviB_time} that at smaller lags the PVI measure captures highly intermittent events, while such events are relatively rare in the time series computed using a 100-second lag. This is reinforced by Figure \ref{fig:pvi_pdf}, which shows histograms of PVI for the three lags. The most probable value of PVI is about 0.5 for all three cases, and corresponds to the majority of events, that are, by definition, non-intermittent. While all three lags capture a large number of non-Gaussian (\(\text{PVI}>3\)) events, the tails of the histograms become wider as the lag is decreased; the 1 and 10-second lags pick out hundreds of current-sheets (\(\text{PVI}>6\)), and possible reconnection sites (\(\text{PVI}>8\)) are detected with 1-second lag.

\begin{figure*}
\centering
\gridline{\fig{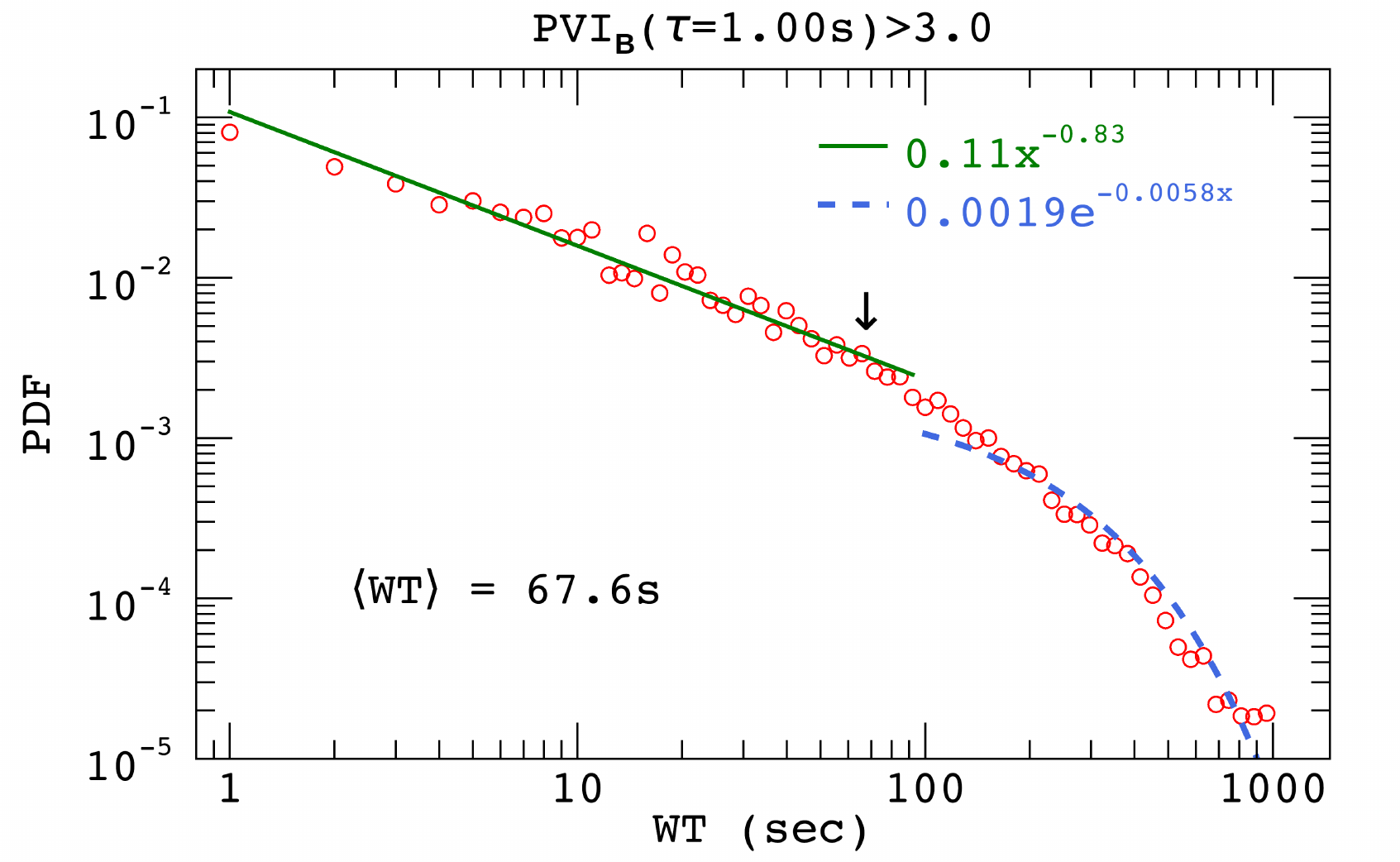}{.5\textwidth}{(a)}
\fig{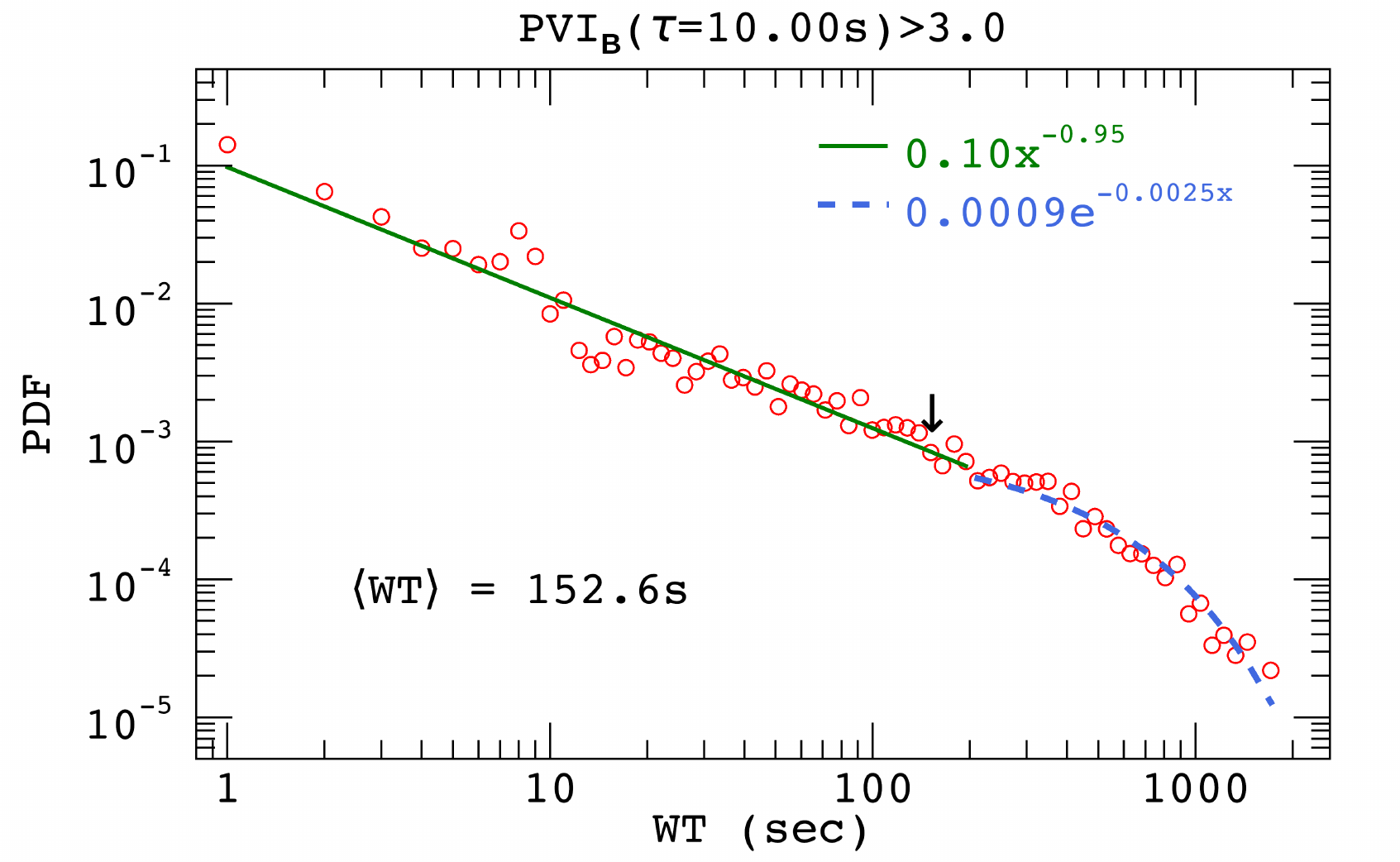}{.5\textwidth}{(b)}}
\gridline{\fig{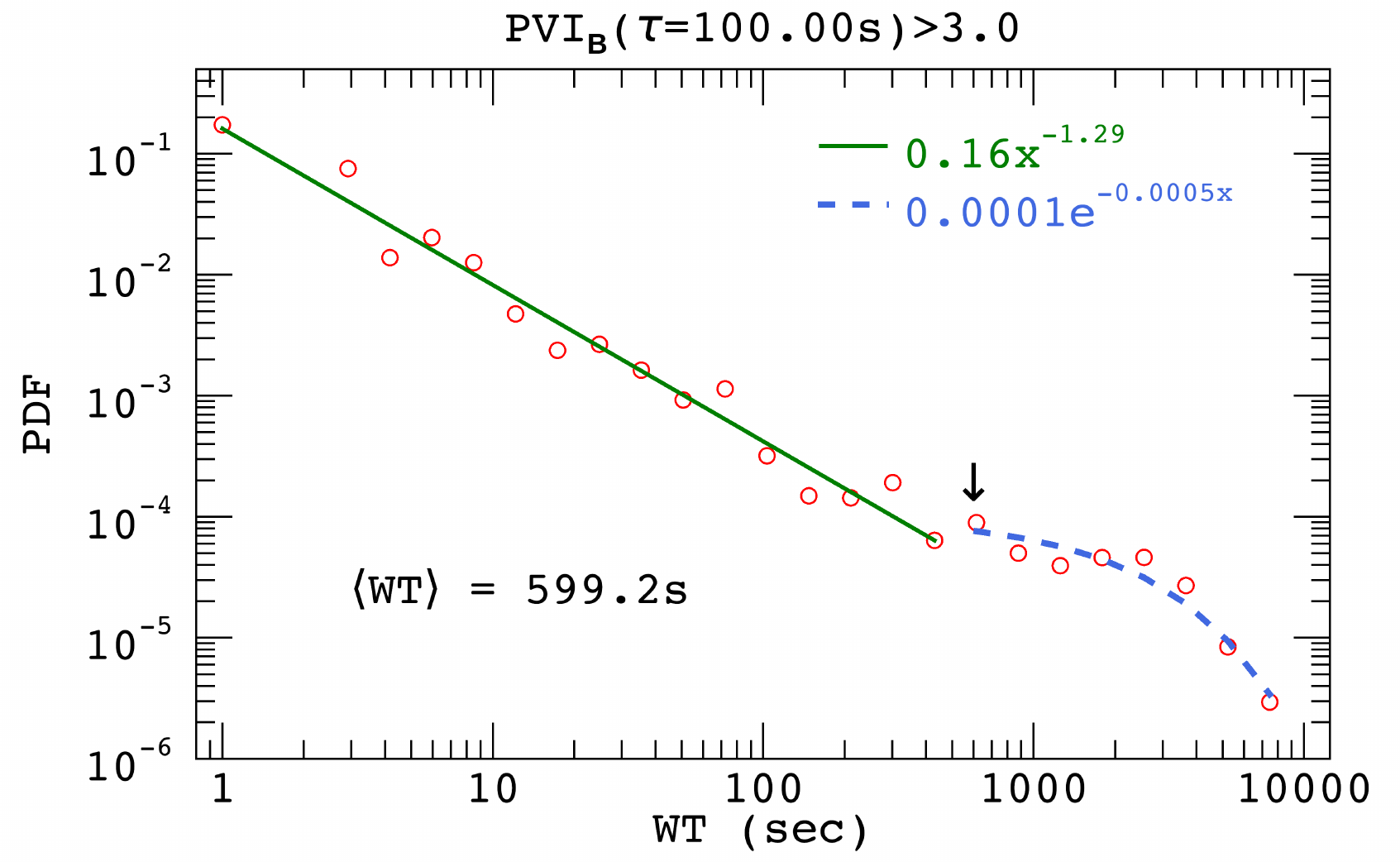}{.5\textwidth}{(c)}
\fig{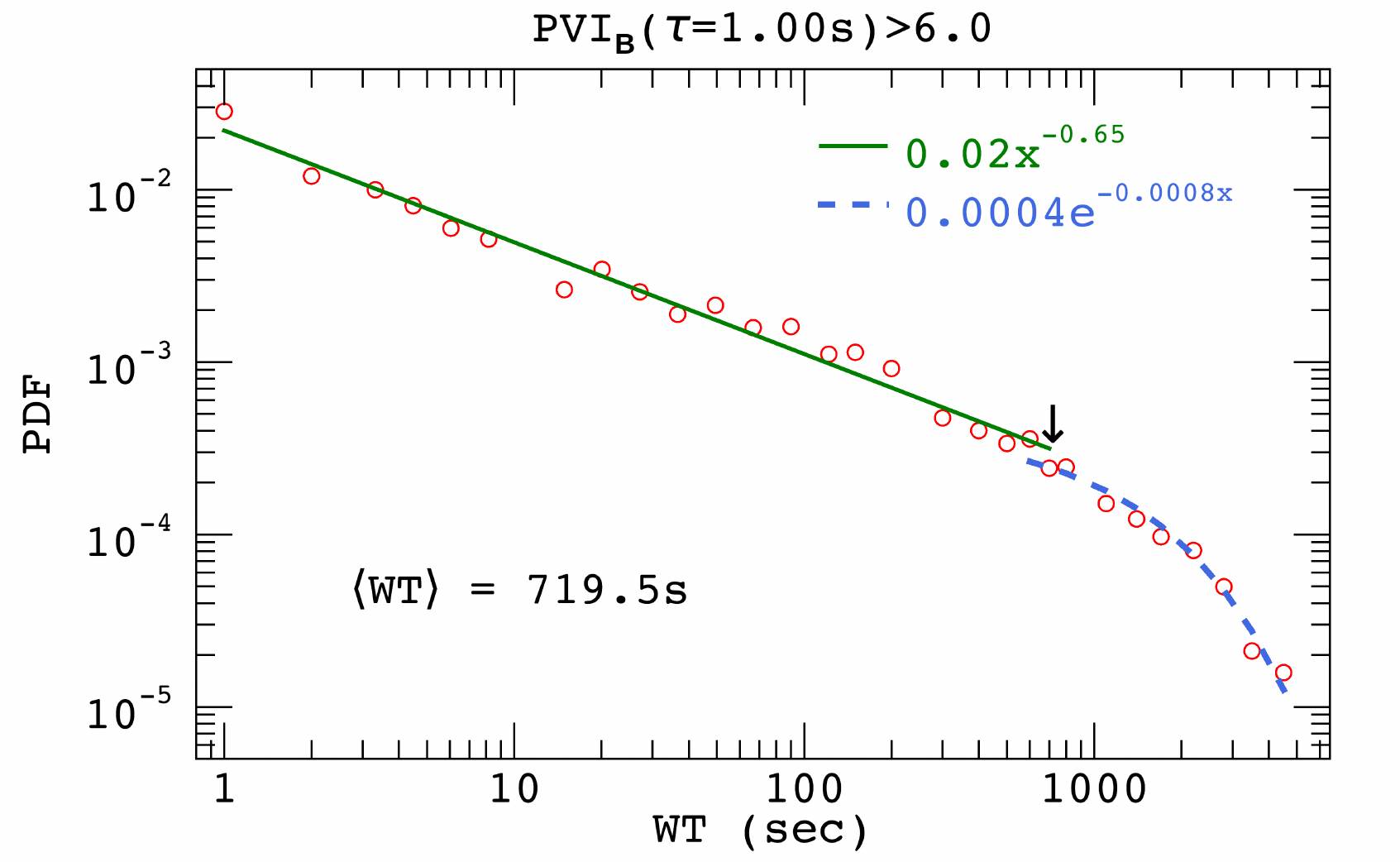}{.5\textwidth}{(d)}
}         
\caption{PDFs of waiting times (WT) between magnetic PVI \(> 3\) events for lags \(\tau\) equal to (a) 1, (b) 10, and (c) 100 seconds. Panel (d) shows the PDF of waiting times between magnetic PVI \(>6\) events for 1-second lag. Bins with fewer than ten counts have been discarded, except in (c) where bins with fewer than 5 counts have been discarded. The average waiting times \(\langle \text{WT}\rangle\) are also indicated, with downward arrows marking their location on the horizontal axes. Power-law fits (\(\beta x^\alpha\)) are shown as solid green lines and exponential fits (\(\gamma e^{-\delta x}\)) as dashed blue curves; here \(x\) refers to the waiting time. 1-sigma uncertainty estimates for fit parameters \(\{\beta, \alpha, \gamma,\delta\}\) for the four panels are, respectively: (a)  \{0.01, 0.03, 0.0002, 0.0003\}; (b) \{0.01, 0.04, 8.8e-05, 0.0001\}; (c)  \{0.04, 0.06, 1.8e-05, 4.8e-05\}; (d) \{0.002, 0.02, 4.8e-05, 4.9e-05\}. The Pearson correlation-coefficients indicating goodness-of-fit for the power-law fits are above 0.95 for each panel.}
\label{fig:WT_pvi3}
\end{figure*}

In Figure \ref{fig:WT_pvi3} we present the main results of this work -- PDFs of waiting times (WT) between intermittent PVI events with varying lag and threshold. Here the waiting time between two events is defined as the time between the end of the first event and the beginning of the second event. Note that the events may have finite duration; i.e., if the PVI stays above the threshold for consecutive times then these times are regarded as part of the same event. Power-law and exponential fits (based on chi-squared error minimization) to the PDFs are also shown, and the average waiting times computed from the distribution are indicated as \(\langle \text{WT}\rangle\). Uncertainties in fit parameters and goodness-of-fit estimates are also reported in the caption.

It is apparent from all four panels of Figure \ref{fig:WT_pvi3} that the distribution of waiting times is well described as a power law for events whose temporal separation is smaller than the correlation time, suggesting strong correlation and clustering. For events that are farther apart in time, the distribution is better fit by an exponential, indicative of a random Poisson-type process \citep{greco2009PRE}. In fact, the break between the power-law and exponential regimes is associated with the average waiting time. While acknowledging that these power-law distributions lack a well-defined average \citep{newman2005ConPh}, we might interpret \(\text{WT} < \langle \text{WT}\rangle\) to be an \textit{intracluster} waiting time, and \(\text{WT} > \langle \text{WT}\rangle\) as an \textit{intercluster} waiting time. The latter is consistent with an exponential, so WT \textit{between} clusters is governed by a uniform random-Poisson process. \textit{Within} clusters, there is strong correlation. \added{We remark here that truncated L\'evy-flight (TLF) distributions include both a power-law range along with an exponential cutoff \citep{bruno2004EPL}, and in future work it would be worth examining the present results in the context of such TLFs.}

Another feature of interest in Figure \ref{fig:WT_pvi3} is the small spike in the PDF near \(\text{WT} = \tau\), suggesting that the PVI at a given lag may preferentially pick out events with a characteristic waiting time equal to the specified lag.\footnote{\footnotesize{This appears to be a general property of the PVI measure that has been seen in previous work \citep[e.g., see Figure 4 in][]{greco2009PRE}, but has either not been noticed or not remarked upon until now. As a tentative explanation, imagine a single data-point with a strong fluctuation relative to its neighbors, and suppose the lag is 100 s. Then, 100 s before this point there is a strong likelihood of a PVI event, and of another PVI event at the time of this point. This could lead to an increased chance of \(\text{WT} = \tau\) compared to neighboring values of WT.}} We also note that the magnitude of the slope $\alpha$ of the power law systematically increases with increasing lag, indicating a weakening of the clusterization (see Appendix  \ref{sec:app}). Furthermore, for \(\tau=1\), the slope is shallower for the case of the PVI \(> 6\) threshold compared to the PVI \(>3\) threshold,
suggesting that more intermittent events are more strongly clustered. Note that we have not examined longer lags for the PVI \(>6\) threshold, since such high PVI values are relatively rare for 10 and 100 second lags (see Figure \ref{fig:pvi_pdf}). We remark here that we also performed this analysis for the periods of 2018 October 17-26 and 2018 November 14-24, and obtained similar results.

In Table \ref{tab:B} we show power law slopes, average waiting times, and average durations of events for the various lags and thresholds. Times have also been converted to distances, assuming Taylor's hypothesis \citep{taylor1938ProcRSL} with a constant average radial speed of 350 \(\text{km}~\text{s}^{-1}\) for the encounter.\footnote{\footnotesize{Since PSP and the solar wind plasma were in near-corotation near perihelion \citep{kasper2019nature}, we reasoned that the plasma was convecting past the spacecraft primarily in the radial direction, and therefore used the radial speed of the solar wind while employing the Taylor hypothesis.}} Note that the Taylor hypothesis has been found to have reasonable validity during the first encounter \citep[see][Chasapis et al. 2019, and Parashar et al. 2019 in the present volume]{Chen2019psp}, consistent with predictions based on turbulence modeling of the solar wind \citep{chhiber2019psp2}. However, the distances shown in Table \ref{tab:B} should be considered crude estimates, since the radial velocity varies by up to a factor of 2 relative to the mean used here. Note that this constant radial speed is used only in estimations of characteristic distances, and plays no role in our temporal analyses and conclusions.

\begin{table*}
  \begin{tabular}{| c | c | c | c | c | c | c |}
    \hline
 & \multicolumn{3}{c|}{\(\text{PVI}_{\bm{B}} >3\)} & \multicolumn{3}{c|}{\(\text{PVI}_{\bm{B}} >6\)} \\ \hline
 \(\tau\) in s (km) &  \(\alpha\) & \(\langle \text{WT}\rangle\)~in s (km) & \(~\langle \text{T}_\text{dur}\rangle\)~in s (km) & \(\alpha\) & \(\langle \text{WT}\rangle\)~in s (km) & \(~\langle \text{T}_\text{dur}\rangle\)~in s (km) \\ \hline
      1 (350) & \(-0.83\) & 67.6 (23,000) & 1.2 (420) & \(-0.65\) & 719.5 (252,000) & 1.0 (350)  \\ \hline
      10 (3500) & \(-0.95\) & 152.6 (53,000) & 2.7 (945) &  &  &  \\ \hline
     100 (35,000) & \(-1.29\) & 599.2 (210,000) & 4.5 (1575) &  &  &  \\ \hline
  \end{tabular}
\caption{Power-law indices \(\alpha\) of fits to WT distributions, mean waiting times \(\langle \text{WT}\rangle\) in s, and mean durations \(\langle \text{T}_\text{dur}\rangle\) in s, for different PVI lags (\(\tau\)) and thresholds, for the magnetic field. Times have been converted to approximate characteristic distances (shown in km in parentheses) assuming Taylor's hypothesis with an average radial speed of 350 km s$^{-1}$. For reference, the mean correlation time (distance) for magnetic fluctuations during the encounter is about 600 s (200,000 km). Above \(\langle\text{WT}\rangle\) the waiting times depart from a power law and follow an exponential distribution.}\label{tab:B}
\end{table*}

\section{PVI Analysis of Velocity}\label{sec:vel}

\begin{figure*}
\centering
\gridline{\fig{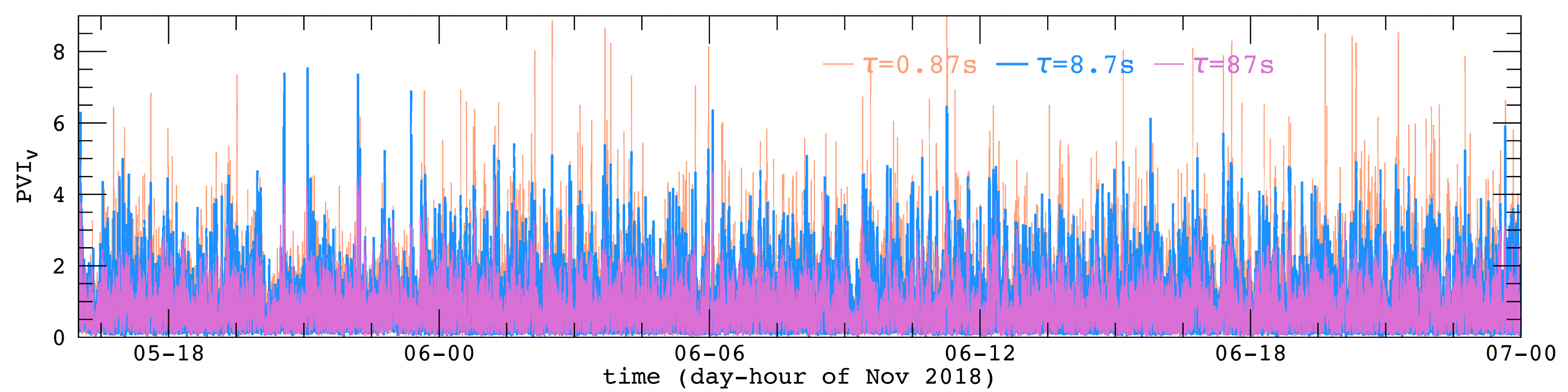}{1\textwidth}{(a)}}
\gridline{\fig{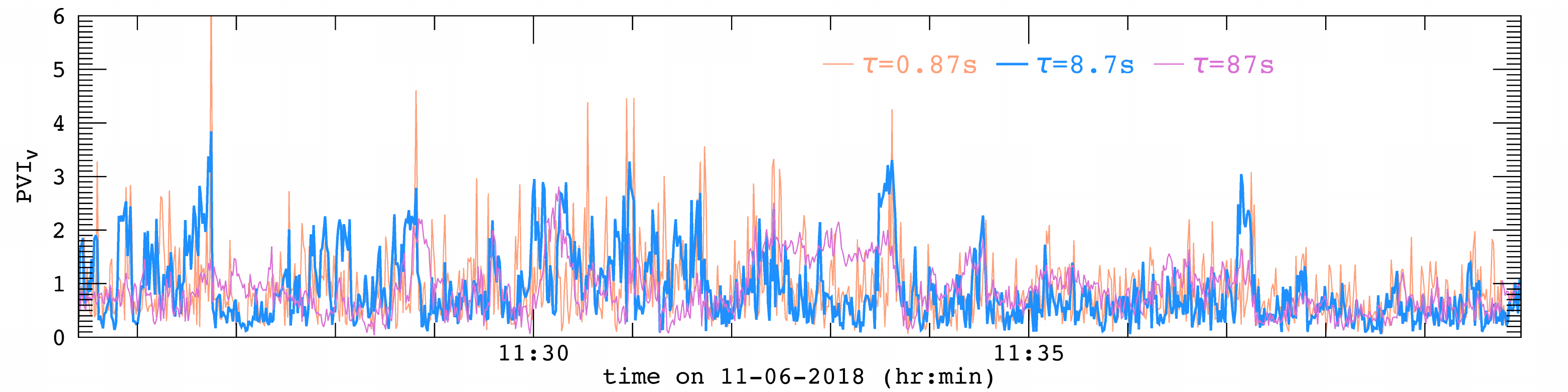}{1\textwidth}{(b)}}
\caption{(a) PVI (with lag \(\tau\) equal to 0.87, 8.7, and 87 s) time-series for the proton velocity from UTC 2018-11-05T16:00:03 to  2018-11-07T00:00:03, including the first perihelion. (b) The same time series for about 15 minutes on 2018 November 6. In both panels the 8.7 s case is shown as a thicker line compared to the other two.}
\label{fig:pviV_time}
\end{figure*}

\begin{figure}
\centering
\includegraphics[width=0.5\textwidth]{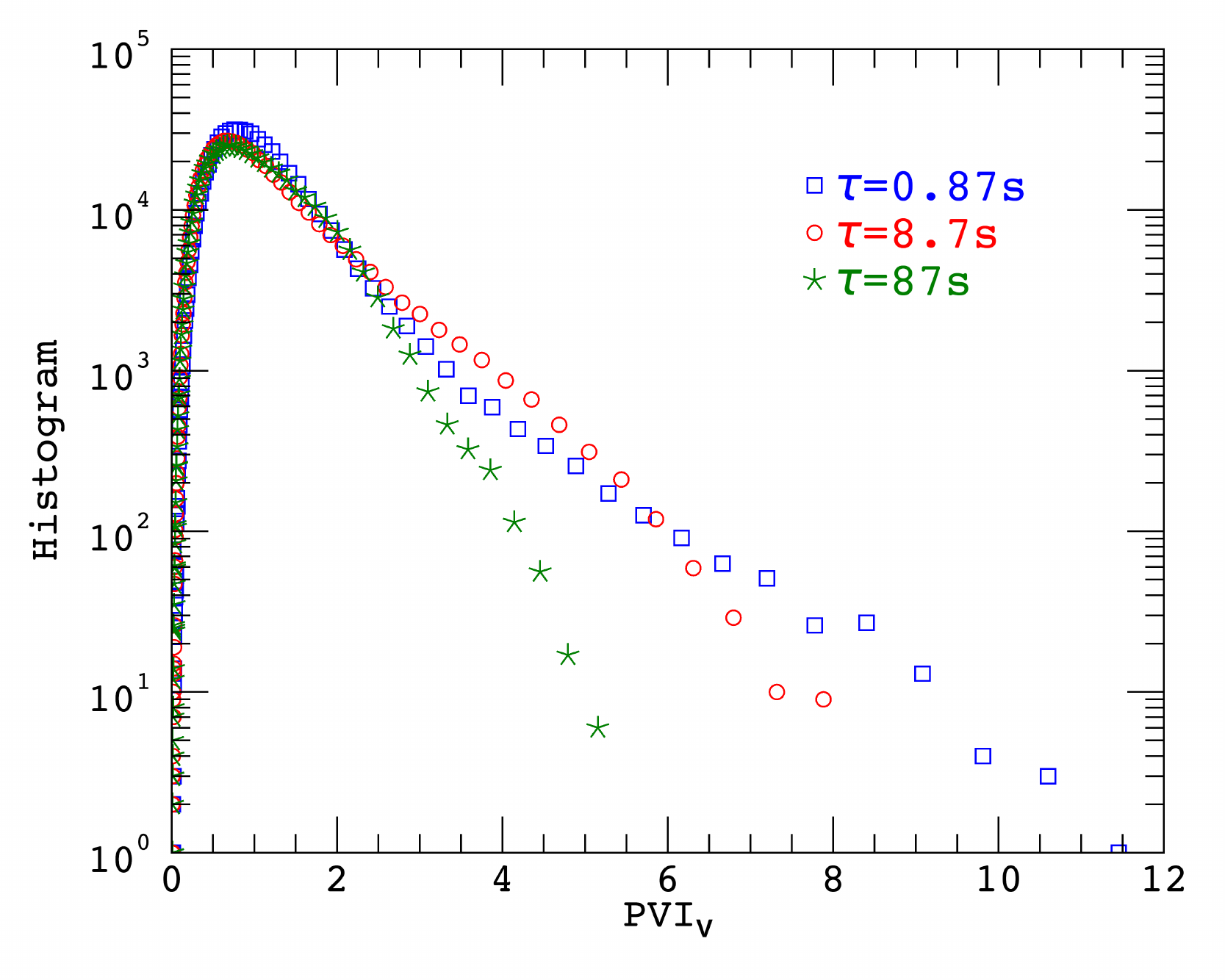}
\caption{Histograms (showing frequency of occurrence, or number of counts) of PVI values for different lags \(\tau\), for the proton velocity during the first encounter. Note the elevated likelihood of large PVI values at shorter lags, indicative of enhanced
small-scale intermittency, typical of non-Gaussian processes and turbulence.}
\label{fig:pvi_pdf_V}
\end{figure}

Next we present the results of the PVI waiting-time analysis for the proton velocity. As discussed in Section \ref{sec:data}, data gaps are a more significant issue for velocity measurements by the SPC, compared with MAG data. Velocity data selection and processing is described in Section \ref{sec:data}. The resulting subsamples (i), (ii), and (iii) have velocity correlation times of 1700, 325, and 700 seconds, respectively, and a rolling average over an interval ten times larger than these times is employed for computing the PVI time-series [Equation \eqref{eq:pvi}]. The time series for the second subsample (near perihelion) is shown in Figure \ref{fig:pviV_time}, for three different lags. As in the case of the magnetic field, smaller lags detect more intermittent events, although there appear to be relatively fewer events with very high PVI. This finding is reinforced by the histograms shown in Figure \ref{fig:pvi_pdf_V}, with the caveat that the volume of data used in the analyses for the velocity is smaller than that for the magnetic field, since in the former case only intervals that survive the data-selection procedure (Section \ref{sec:data}) are used. Nevertheless, we do find thousands of non-Gaussian events (PVI \(>3\)) and more than a hundred possible current sheets (PVI \(>6\)).


\begin{figure*}
\centering
\gridline{\fig{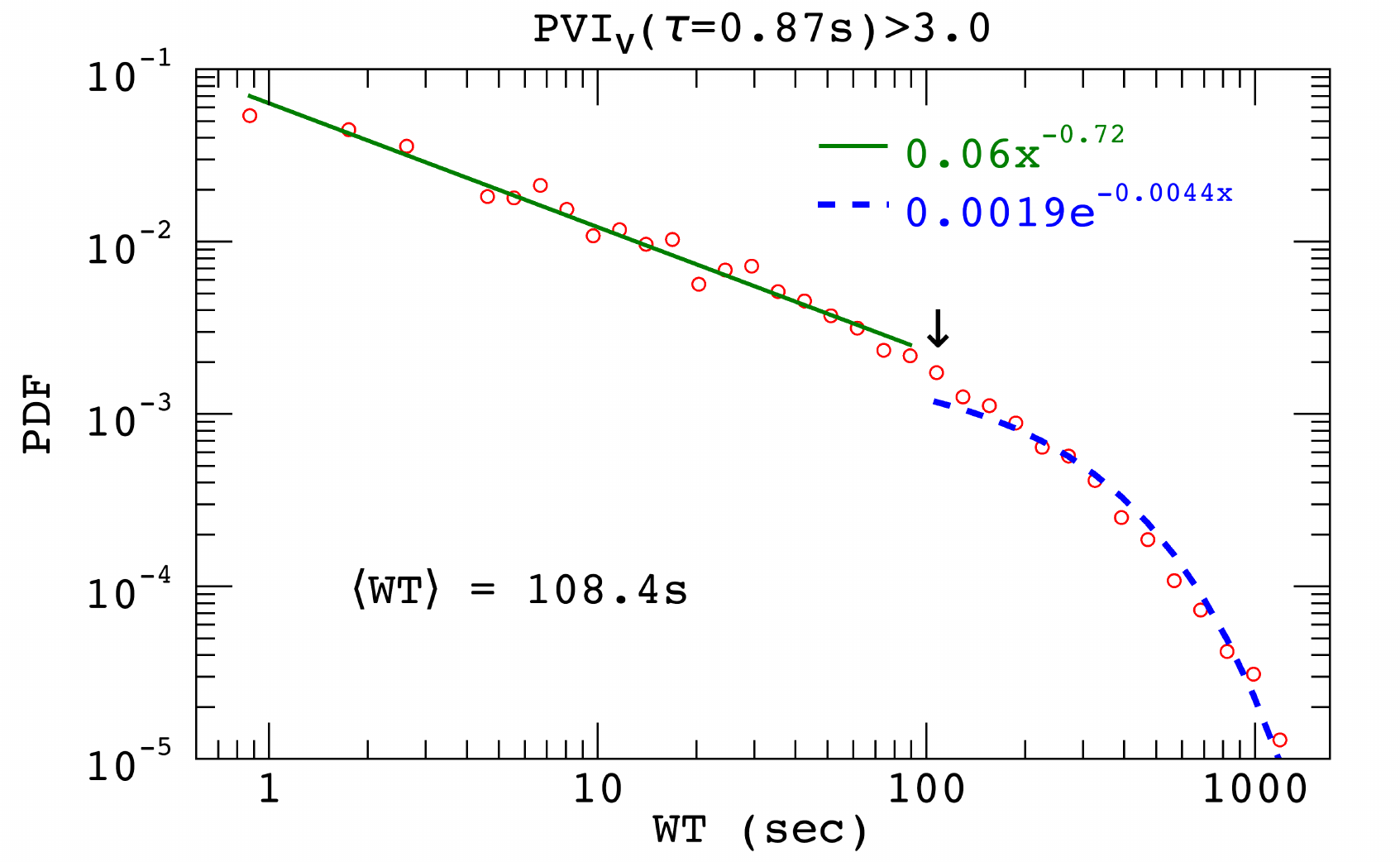}{.5\textwidth}{(a)}
\fig{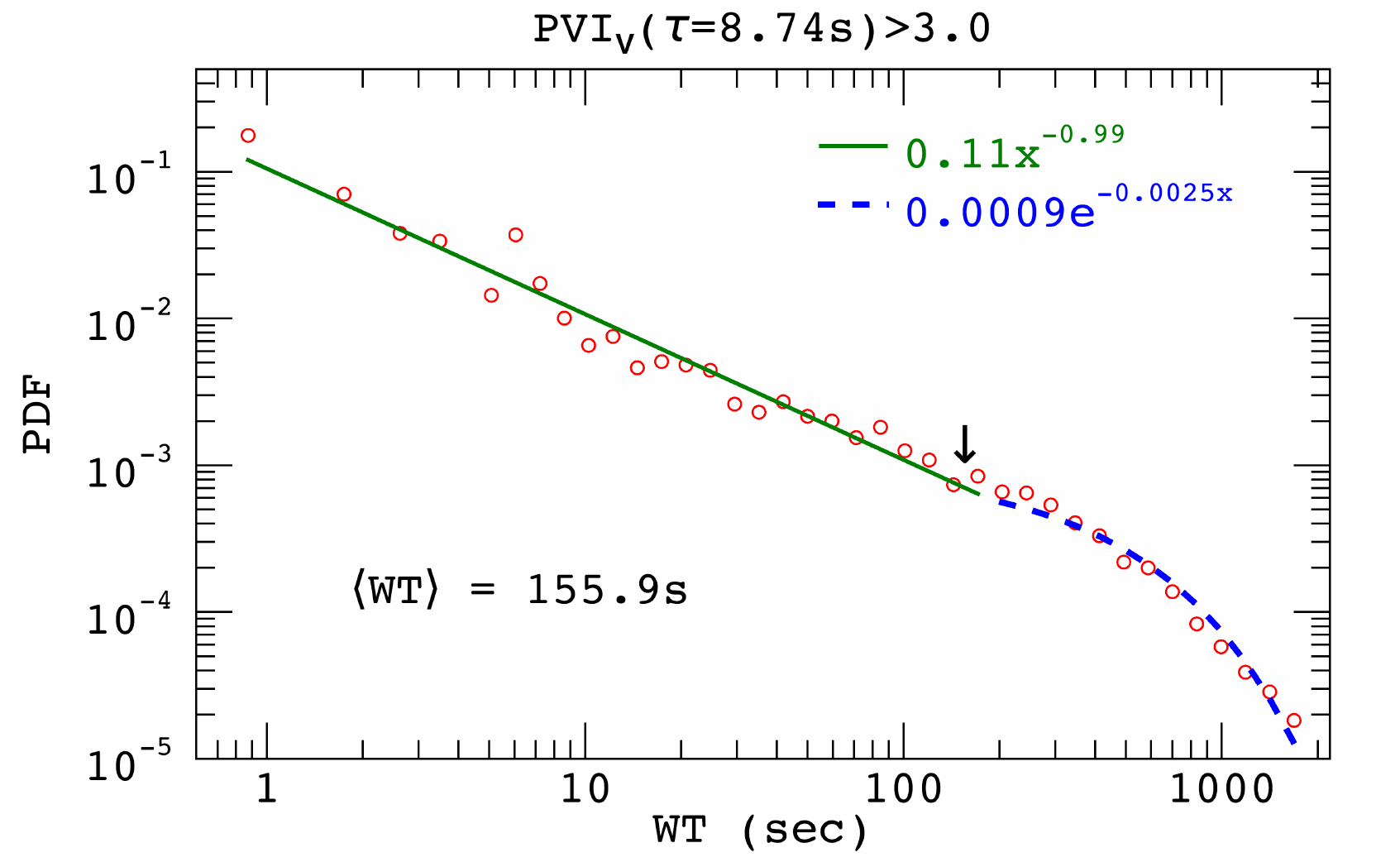}{.5\textwidth}{(b)}}
\gridline{\fig{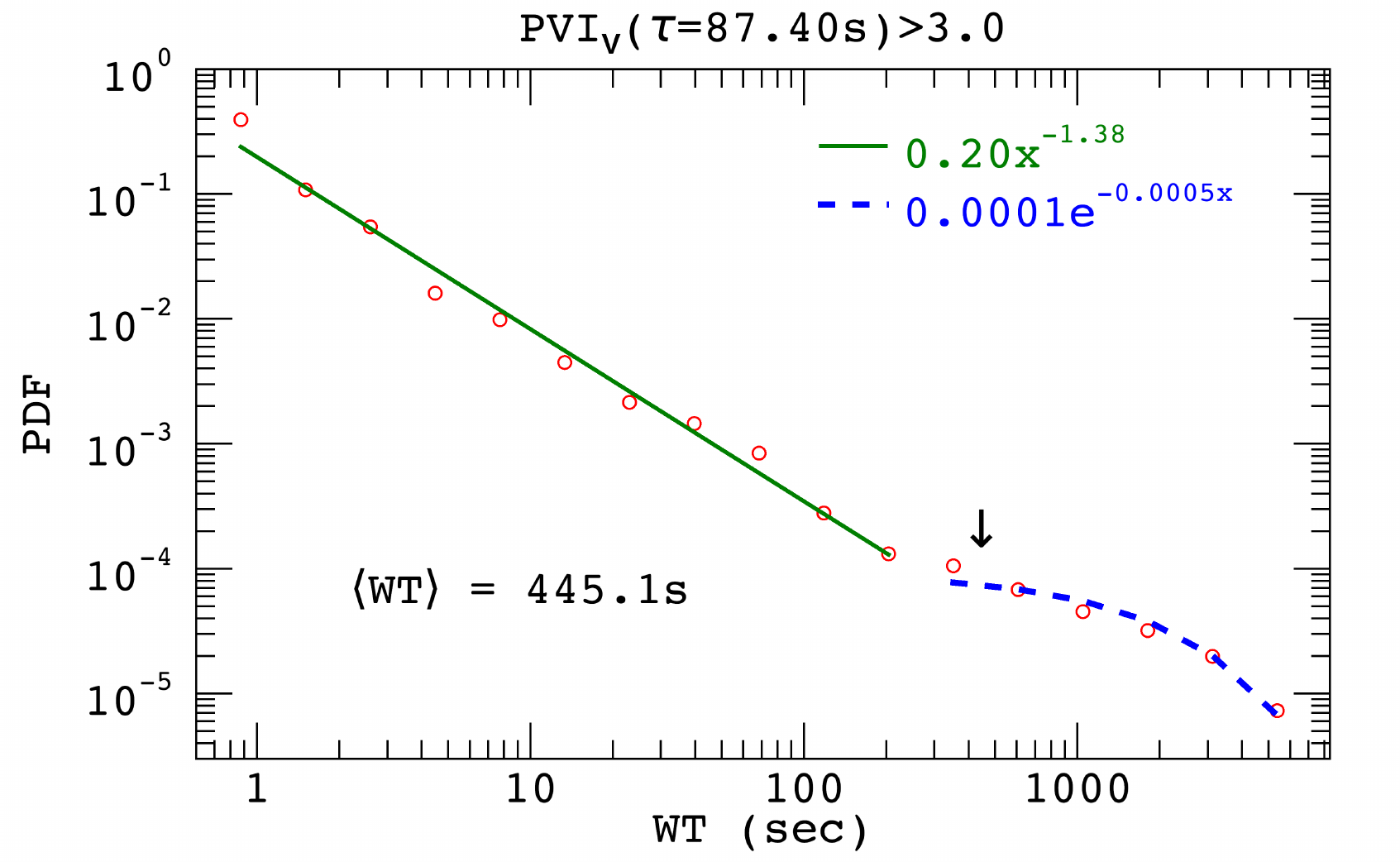}{.5\textwidth}{(c)}
\fig{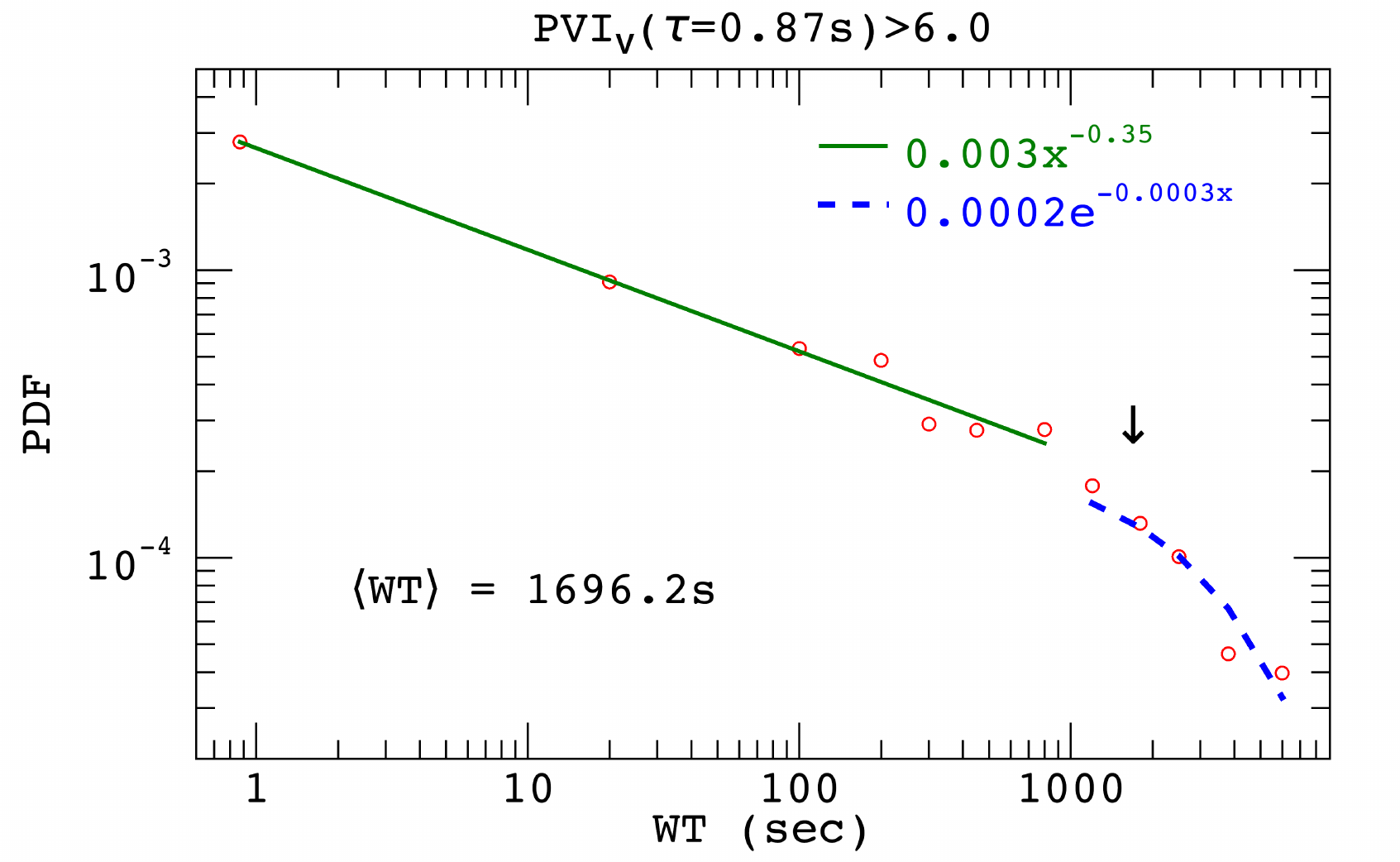}{.5\textwidth}{(d)}
}         
\caption{PDFs of waiting times between (proton) velocity PVI \(> 3\) events for PVI lags (a) 0.87, (b) 8.7, and (c) 87 seconds. Panel (d) shows the PDF of waiting times between (proton) velocity PVI \(>6\) events for 0.87 second lag. Bins with fewer than ten counts have been discarded in panels (a) to (c), while in panel (d) bins with fewer than five counts have been discarded. The average waiting times \(\langle \text{WT}\rangle\) are also indicated, with downward arrows marking their location on the horizontal axes. Power-law fits (\(\beta x^\alpha\)) are shown as solid green lines and exponential fits (\(\gamma e^{-\delta x}\)) as dashed blue curves; here \(x\) refers to the waiting time. 1-sigma uncertainty estimates for fit parameters \(\{\beta, \alpha, \gamma,\delta\}\) for the four panels are, respectively:  (a) \{0.005, 0.03, 0.0002, 0.0002\}; (b) \{0.01, 0.04, 0.0001, 0.0001\}; (c) \{0.03, 0.05, 1.2e-05, 4.9e-05\}; (d) \{0.0003, 0.02, 5.3e-05, 6.6e-05\}. The Pearson correlation-coefficients indicating goodness-of-fit for the power-law fits are above 0.95 for each panel.} 
\label{fig:WT_pvi3_V}
\end{figure*}

Moving on to distributions of waiting times between velocity PVI events, the results in Figure \ref{fig:WT_pvi3_V} are consistent with the magnetic case (Figure \ref{fig:WT_pvi3}). The PDFs are described well by power laws up to the average waiting time, and are fit better by exponentials for larger waiting times. Once again this behavior suggests strong correlations within clusters of size \(\langle \text{WT}\rangle\), and random Poisson intercluster processes. The magnitude of the slope \(\alpha\) of the power law increases with increasing lag, and, for the 0.87 s lag, the slope of the PVI \(>6\) power law is shallower than the PVI \(>3\) case, suggesting that stronger intermittency is associated with increased clusterization. The small spike in the PDF at \(\text{WT}=\tau\) is also seen here.

Table \ref{tab:V} shows power-law slopes, average waiting times, and average durations of PVI events for the various lags and thresholds considered. Times have been converted to approximate characteristic distances assuming Taylor's hypothesis with a constant speed of 350 \(\text{km}~\text{s}^{-1}\), as in the magnetic case.

\begin{table*}
\begin{center}
  \begin{tabular}{| c | c | c | c | c | c | c |}
    \hline
 & \multicolumn{3}{c|}{\(\text{PVI}_{\bm{V}} >3\)} & \multicolumn{3}{c|}{\(\text{PVI}_{\bm{V}} >6\)} \\ \hline
 \(\tau\) in s (km) &  \(\alpha\) & \(\langle \text{WT}\rangle\)~in s (km) & \(~\langle \text{T}_\text{dur}\rangle\)~in s (km) & \(\alpha\) & \(\langle \text{WT}\rangle\)~in s (km) & \(~\langle \text{T}_\text{dur}\rangle\)~in s (km) \\ \hline
      0.87 (305) & \(-0.72\) & 108.4 (38,000) & 0.98 (343)  & \(-0.35\) & 1696.2 (594,000) & 0.89 (312)  \\ \hline
      8.7 (3045) & \(-0.99\) & 155.9 (55,000) & 2.3 (805) &  &  &  \\ \hline
     87 (30,450) & \(-1.38\) & 445.1 (156,000) & 3.2 (1120) &  &  &  \\ \hline
  \end{tabular}
\caption{Power law indices \(\alpha\) of fit to WT distributions, mean waiting times \(\langle \text{WT}\rangle\) in s, and mean durations \(\langle \text{T}_\text{dur}\rangle\) in s, for different PVI lags (\(\tau\)) and thresholds, for the proton velocity. Times have been converted to distances (shown in km in parentheses) assuming Taylor's hypothesis with an average radial speed of 350 \(\text{km}~\text{s}^{-1}\). For reference, the mean correlation time (distance) for velocity fluctuations near perihelion is about 325 s (114,000 km). Above \(\langle\text{WT}\rangle\) the waiting times depart from a power law and follow an exponential distribution.}\label{tab:V}
\end{center}
\end{table*}

\section{Conclusions and Discussion}\label{sec:conclude}

In this paper we have employed the PVI methodology to 
provide a baseline statistical characterization 
of the ``roughness'', or intermittency, of the observed 
magnetic and velocity field during the first solar encounter of the PSP. Quantification of roughness using the PVI technique has the dual advantages of being 
closely related to turbulence intermittency 
statistics, while also being 
related to classical-discontinuity identification
procedures \citep{greco2009PRE,greco2018ssr}.\footnote{\footnotesize{By ``classical discontinuity'' we are referring to the traditional interpretation of (mostly magnetic) discontinuities in the solar wind as members of a class of MHD stationary convected structures (such as tangential discontinuities) or propagating rotational discontinuities, which are viewed as static solutions of the ideal MHD equations \citep[e.g.,][]{neugebauer2010SWconf}.}}
The present work extends in a natural way analogous studies 
carried out at 1 au and beyond \citep{greco2018ssr}.
Values of PVI above appropriate thresholds have been found to be related to classical discontinuities \citep{greco2008GRL,greco2009PRE}, 
intermittency and current sheets \citep{greco2009ApJ,malaspina2013JGR}, particle energization \citep{tessein2013ApJ,tessein2015apj,tessein2016GRL}, kinetic effects such as 
elevated temperature and high degrees of non-Guassianity in the velocity distribution 
function \citep{osman2011ApJ,osman2012PRL,servidio2015JPP,Qudsi2019psp}, and, at high PVI, likelihood of 
magnetic reconnection \citep{servidio2011JGR}. 
In this sense it is a natural follow-on to 
examine whether those tendencies extend
further into the inner heliosphere than has been previously explored.
However, additional motivation is obtained through early reports that the magnetic and velocity fields near PSP perihelion
exhibit strong ``jets'' or ``switchbacks''
that may suggest enhanced, episodic, and 
large-amplitude quasi-discontinuous jumps in the 
plasma conditions (see several papers in this special edition). PVI seems to be an appropriate general tool for broadly identifying and quantifying such intermittent structure. Note that further detailed study of specific types of structures, such as the observed ``switchbacks'', requires a more specialized approach (see Dudok de Wit et al. 2019, current issue).

Our main results are
summarized in the Tables.
During the first Parker Solar Probe 
encounter, the fluctuations of 
both the magnetic 
field and velocity field
exhibit statistical features, specifically the 
inter-event waiting-time distributions, that 
suggest the appearance of both correlated  as well as 
random or Poissonian events. Such events are interpreted as non-Gaussian coherent structures, consistent with current sheets and vortex sheets.
The presence of these signals 
may be related 
to interpretations
based on 
intermittent turbulence, although 
the method itself is also sensitive to classical discontinuities.
For waiting times shorter than about a correlation scale, 
the presence of power-law distributions indicates correlations
and is suggestive of clustering.
In Appendix \ref{sec:app} we consider an analogy with generalized self-similar Cantor sets, for which the power-law index $\alpha$ ranges from \(-2\) to \(-1\) with increasing clustering. 
For waiting times larger 
than about the measured correlation
scales, the exponential distribution of waiting times indicates uncorrelated random events. The clustering appears to weaken with increasing PVI lag, and, for the same lag, more intermittent events are more strongly clustered. The behavior is consistently seen in both magnetic and velocity fields; this is perhaps not surprising, given the high Alfv\'enicity of the fluctuations observed during the encounter \citep[][Parashar et al. 2019, present volume]{bale2019nature,Chen2019psp}. Our results complement those of Dudok de Wit et al. (2019, present issue), who find that waiting times between the observed ``switchbacks'' in the radial magnetic field are well-described by power-laws. 

These findings appear to be consistent with some recent studies of near-Earth solar-wind fluctuations \citep{greco2009PRE}, as well as simulations of MHD turbulence \citep{greco2008GRL}. Note that the correlation time increases as one moves outward towards 1 au \citep{breech2008turbulence,bruno2013LRSP,Ruiz2014SoPh,Zank2017ApJ835} and the turbulence ``ages'' \citep{matthaeus1998JGR}; therefore one expects the shift from power-law to exponential behavior to occur at larger waiting times compared to the present results for the young solar wind. Indeed, here we find average waiting times of about 3 -- 10 minutes, which are smaller than the typical values of 30 -- 50 minutes seen at 1 au \citep{tsurutani1979JGR,bruno2001PlanSS,greco2008GRL}. Note that other studies have found exponential waiting-time distributions for intermittent events in the near-Earth (and beyond) solar wind \citep{tsurutani1979JGR,bruno2001PlanSS}, without a power-law regime. Interestingly, \cite{hu2018ApJS} find power-law distributions at \textit{longer} waiting times (\(>60\) minutes) and exponential behavior before that, for small-scale flux ropes identified using a Grad-Shafranov reconstruction technique with \textit{WIND} observations. 

\added{Assuming wind speed as the sole criterion for classification, the current PSP observations are mostly restricted to slow-wind conditions in the ecliptic during solar minimum. Future orbits are expected to sample extended periods of fast wind as well, and it will be interesting to compare waiting-time statistics between slow and fast wind in the near-Sun plasma. Farther away, \textit{Helios} observations find power-law behavior up to longer waiting times in the case of slow wind compared with fast wind \citep{DAmicis2006AnGeo}.} It would also be interesting to use full-cadence MAG data (or search-coil magnetometer measurements) from PSP to probe PVI events at kinetic-scale lags.

The dichotomy betweeen a strongly-correlated clustering process and a random Poissonian process may be related to two contrasting views of the origin of magnetic structures in the solar wind -- in-situ generation via turbulent cascade vs. passive advection from the solar source. The strong clustering seen in our present results readily leads to the suggestion that these observed features may originate in a hierarchy of  nonlinear processes that generate correlations of nearby structures over a broad range of scales. Our preferred explanation is strong turbulence occurring in the corona and/or interplanetary medium. Turbulence is known to produce features of the type reported here, as has been observed routinely  
in space plasmas including the solar wind and the terrestrial magnetosheath \citep{yordanova2008prl,matthaeus2011SSR,bruno2013LRSP}. The unique feature of the present analysis is finding these indicators of intermittency and turbulence at distances closer to the Sun, and therefore closer to source and boundary surfaces,
than has been accomplished in any previous space mission. This may eventually
produce constraints on how turbulence is initiated in the inner heliosphere, or how it is transmitted and propagated from the corona into the super-Alfv\'enic solar wind. Fully satisfactory  answers to such questions will likely require additional 
complementary observations by PSP in subsequent orbits, and by the upcoming Solar Orbiter mission. Furthermore, it is likely that more complete interpretations will require context support from global heliospheric simulations to establish likely connections between in-situ observation and remote sensing of the inner solar atmosphere, for example by Solar Orbiter or by the upcoming PUNCH mission.

\section {Acknowledgments}
 This research as been supported 
 in part by the Parker Solar Probe mission under the 
 \isois~project 
 (contract NNN06AA01C) and a subcontract 
 to University of Delaware from
 Princeton (SUB0000165).
 Additional support is acknowledged from the  NASA LWS program  (NNX17AB79G) and the HSR program 
 (80NSSC18K1210 \& 80NSSC18K1648), and grant RTA6280002 from Thailand Science Research and Innovation. Parker Solar Probe was designed, built, and is now operated by the Johns Hopkins Applied Physics Laborotary as part of NASA's Living With a Star (LWS) program (contract NNN06AA01C). Support from LWS management and technical team has played a critical role in the success of the Parker Solar Probe mission.
 
\appendix

\section{Waiting Times for the Cantor Set}\label{sec:app}
Here we provide details of the 
association between
power-law waiting times 
and processes or structures that can 
be described by a Cantor set.

For a given power-law distribution of waiting times, with ${\rm PDF}\propto{\rm WT}^{\alpha}$, it may not be clear how to physically interpret the power-law index $\alpha$.  Intuitively, it seems that a harder distribution should indicate stronger clustering than a softer distribution, i.e., $\alpha\approx-1$ should indicate stronger clustering than $\alpha\approx-2$, because a process with a harder waiting time distribution more frequently has a long hiatus followed by numerous events in rapid succession.  

To interpret $\alpha$ more quantitatively, and given that (statistical) self-similarity is a common feature of inertial-range turbulence, we consider the waiting-time distribution of the Cantor set \citep{smith1874ProcLondMathSoc,cantor1883}.  Recall that this set is defined as the points remaining after an infinite sequence of operations: At stage $n=0$ we start with the set $[0,1]$, then in stage $n=1$ we remove the middle 1/3 with 2 segments remaining at either side, and in each subsequent stage $n$ we remove the middle 1/3 of each remaining segment, doubling the number of remaining segments to become $2^n$.  If the ``waiting time'' $T$ is defined as the distance between successive points in the Cantor set, then all waiting times are $T_n=3^{-n}$ for some $n\in\{1,2,3,\dots\}$, and the number of waiting times generated in stage $n$ is $N_n=2^{n-1}$.  An unnormalized PDF of waiting times can be defined as $N_n/(T_n-T_{n+1})$, which results in 
\begin{equation}
{\rm PDF}(T_n=3^{-n})=\frac{2^{n-1}}{(2/3)3^{-n}}=\frac{9}{2}6^{n-1}.  
\end{equation}
This implies that 
\begin{equation}
\alpha = \frac{\ln {\rm PDF}(T_{n+1})-\ln {\rm PDF}(T_n)}{T_{n+1}-T_n} = -\frac{\ln6}{\ln3}\approx\ -1.631
\end{equation}
[The same power-law index results if we instead define the PDF from $N_n/(T_{n-1}-T_n)$.]  Remarkably, some of the present observational results for PVI events have $\alpha$ close to -1, implying that large field-increments in the solar wind can be more strongly clustered than the Cantor set. Similar slopes have been observed near 1 au \citep{greco2009PRE}.

As a generalization of the Cantor set, consider a set in which at each stage, instead of removing 1/3 of each segment, we remove a fraction $f$ of the segment from the middle.  As $f\to1$, more of the segment is removed and the remaining points are more clustered with wider gaps.  Each remaining segment after $n=1$ has a size $(1-f)/2$, and after stage $n$ the segment size is $[(1-f)/2]^n$.  Then $T_n=f[(1-f)/2]^{n-1}$ and we still have $N_n=2^{n-1}$, so 
\begin{equation}
{\rm PDF}(T_n)=\frac{2^{n-1}}{f[(1+f)/2][(1-f)/2]^{n-1}} = \frac{2}{f(1+f)}\left(\frac{4}{1-f}\right)^{n-1}
\end{equation}
and 
\begin{equation}
\alpha = \frac{\ln[4/(1-f)]}{\ln[(1-f)/2]} = -\frac{2\ln2-\ln(1-f)}{\ln2-\ln(1-f)}.
\end{equation}
For $0<f<1$, we have $-2<\alpha<-1$, with $\alpha\to-1$ as $f\to1$.  Thus our observation of $\alpha\approx-1$ for field increments in the solar wind implies extreme clustering, equivalent to the maximum clustering possible for such a generalized Cantor set.

\bibliography{chhibref}

\listofchanges
\end{document}